\documentclass[aps,prl,twocolumn,raggedbottom,showpacs,superscriptaddress,10pt]{revtex4-1}

\usepackage{graphicx}
\usepackage{color}
\usepackage{amsmath}
\usepackage{amsfonts}
\usepackage{bm}
\usepackage{enumerate}
\usepackage{array}

\newcommand{\vect}[1]{\mathbf{#1}}
\newcommand{\bra}[1]{\ensuremath{\langle #1|}}
\newcommand{\ket}[1]{\ensuremath{|#1\rangle}}
\newcommand{\braket}[2]{\ensuremath{\langle #1|#2\rangle}}
\newcommand{\rmd}{\mathrm{d}}
\newcommand{\rme}{\mathrm{e}}
\newcommand{\rmi}{\mathrm{i}}
\newcommand{\Tr}{\mathrm{Tr}}

\begin{document}

\title{Seeing bulk topological properties of band insulators in small photonic lattices}

\author{Charles-Edouard Bardyn}
\affiliation{Institute for Quantum Electronics, ETH Zurich, 8093 Z{\"u}rich, Switzerland}
\author{Sebastian D. Huber}
\affiliation{Institute for Theoretical Physics, ETH Zurich, 8093 Z{\"u}rich, Switzerland}
\author{Oded Zilberberg}
\affiliation{Institute for Theoretical Physics, ETH Zurich, 8093 Z{\"u}rich, Switzerland}

\begin{abstract}

We present a general scheme for measuring the bulk properties of non-interacting tight-binding models realized in arrays of coupled photonic cavities. Specifically, we propose to implement a single unit cell of the targeted model with tunable twisted boundary conditions in order to simulate large systems and, most importantly, to access bulk topological properties experimentally. We illustrate our method by demonstrating how to measure topological invariants in a two-dimensional quantum Hall-like model.

\end{abstract}


\maketitle



In recent years, advances in the engineering of photonic crystals~\cite{Lahini09,Kraus12_1,Verbin13,Rechtsman13,Hafezi13_1,Jia13} and superconducting circuits~\cite{Houck12,Underwood12,Schmidt13} have made it possible to simulate fascinating aspects of condensed matter physics. This bottom-up approach to quantum simulation, however, faces its own challenges since photons---as opposed to electrons---typically weakly interact and, most crucially, do not exhibit fermionic statistics. The technological difficulty of suppressing disorder in large arrays of photonic cavities as well as the unavoidable presence of photon losses additionally constrains the scalability of such systems, rendering the simulation of bulk properties particularly challenging.

Of recent interest in condensed matter physics are bulk \emph{topological} phenomena. In the field of topological phases of matter, one typically deals with gapped systems such as band insulators and superconductors~\cite{Hasan10,Qi11}, in which energy bands are characterized by non-trivial topological invariants, leading to remarkable observable phenomena. While the study of topological phases originated in condensed matter physics, recent experiments based on photonic crystals~\cite{Kraus12_1,Rechtsman13,Kraus12_2,Hafezi13_1} have opened up a similar avenue for research in photonic systems.

Bulk topological properties, which are properties of energy bands \emph{as a whole}, are only expected to manifest through observable physical quantities when bands are populated \emph{uniformly}. In fermionic systems, energy bands fill up naturally owing to the Pauli exclusion principle, leading to a variety of observable topological responses such as quantized Hall conductances in two-dimensional (2D) electron gases~\cite{Klitzing80} and quantized magnetoelectric effects in 3D topological insulators~\cite{Qi08}. In these cases, fermionic statistics is key to the observation of bulk topological properties, raising the intriguing question of how to observe such properties in bosonic systems and in photonic ones, in particular.

In photonic systems, the absence of any exclusion principle and unavoidable photon losses make it particularly challenging to populate energy bands uniformly. Although entire bands can in principle be addressed using a driving field of suitable bandwidth, the fact that the injection of photons additionally depends on the spatial overlap between the driving field and the eigenstates of the targeted band invariably leads to non-uniform filling. The issue of measuring topological invariants in photonic lattices was partly addressed in two recent proposals using edge states~\cite{Hafezi13_2} and approximate responses~\cite{Ozawa13}.

\begin{figure}[t]
    \begin{center}
        \includegraphics[width=\columnwidth]{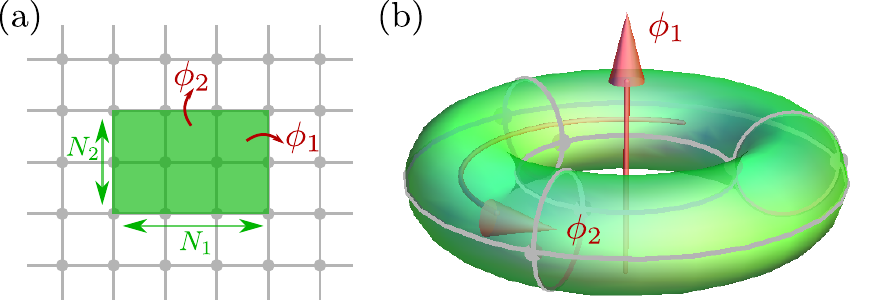}
        \caption{(Color online) \textit{Quantum simulation in the unit cell.} (a) Unit-cell configuration consisting of $N_u$ lattice sites in each spatial direction $u$ (here $u = 1, 2$ in 2D), with twisted boundary conditions defined by periodic boundary conditions with additional twist angles $\phi_u$ at the boundaries. (b) Gauge-equivalent configuration with periodic boundary conditions and fluxes $\phi_u$ threading the unit cell.}
        \label{fig:unitCellSimulator}
    \end{center}
\end{figure}

In this Letter, we present a robust and practical method to measure the bulk topological properties of photonic systems, thus allowing to simulate the physics of band insulators in the latter. Our approach crucially relies on the realization of a small (unit-cell-sized) photonic lattice system with tunable twisted boundary conditions. On one hand, the restriction to a small system provides a way to circumvent the issue of non-uniform band fillings. On the other hand, the use of tunable twisted boundary conditions restores our ability to study larger systems with translation invariance~\cite{Thouless82,Avron83}. We illustrate the power and limitations of our scheme by investigating in detail the 2D quantum Hall-like Hofstadter model~\cite{Hofstadter76}, and demonstrate that the topological invariant (Chern number) characterizing the bands of such a model can be measured exactly.


\emph{Unit-cell photonic quantum simulation.}---We consider a generic non-interacting tight-binding (lattice) model
\begin{align} \label{eq:genericHamiltonian}
    H = \sum_{i,j} a^\dagger_i \mathcal{H}_{ij} a_j,
\end{align}
where $a^\dagger_i$ ($a_i$) are operators creating (annihilating) particles on a lattice site $i$ and $\mathcal{H}$ is an arbitrary first-quantized single-particle Hamiltonian. We wish to study the bulk properties of this model by simulating it on a lattice of photonic cavities coupled via photon tunneling. To this end, we assume that each lattice site $i$ corresponds to a cavity supporting a single mode with resonance frequency $\mathcal{H}_{ii} = \omega_i$, and that photon tunneling occurs between distinct cavities $i$ and $j$ with amplitude $\mathcal{H}_{ij} = t_{ij} \rme^{\rmi \phi_{ij}}$, where $t_{ij} > 0$ and $\phi_{ij} \in [0, 2\pi)$.

In practice, such a photonic simulator does not perfectly capture the Hamiltonian physics of interest. As in any photonic system, photon losses must be taken into account and compensated for by a drive. Assuming that photons are injected into the system using a monochromatic coherent driving field with frequency $\Omega$, the relevant dynamics takes the form of a standard master equation in Lindblad form
\begin{align} \label{eq:masterEquation}
    \resizebox{.9\hsize}{!}{$\displaystyle \partial_t \rho = -\rmi \left[ H + H_\text{drive}, \rho \right] + \gamma \sum_i \left( 2 a_i \rho \, a_i^\dagger - \{ a^\dagger_i  a_i, \rho \} \right),$}
\end{align}
where $\rho$ is the density matrix of the system, $\gamma$ is the decay rate of the cavities (assumed to be uniform, for simplicity), $H_\text{drive} = \sum_i (f_{i} \rme^{-\rmi \Omega t} a_i + \text{h.c.})$ is the Hamiltonian of the drive, where $f_i$ is the amplitude of the driving field at the location of the $i^\mathrm{th}$ cavity, and we have set $\hbar = 1$.

Starting from empty cavities, the driven-dissipative dynamics described by Eq.~\eqref{eq:masterEquation} leads to a state which is, at all times, a product of local coherent states. More specifically, each cavity $i$ is found in a coherent state determined by a single complex parameter $\alpha_i \equiv \Tr (\rho \, a_i)$ whose evolution is governed by the classical equation
\begin{align} \label{eq:classicalFieldEvolution}
    \rmi \partial_t \ket{\alpha} = (\mathcal{H} - (\Omega + \rmi \gamma)) \ket{\alpha} + \ket{f},
\end{align}
where we have introduced the Dirac vector notation $\ket{x} = (x_1, x_2, \ldots, x_N)$ and have moved to a rotating frame with frequency $\Omega$ by the transformation $\rme^{-\rmi \Omega t} \alpha_i \to \alpha_i$. In steady state, the coherent cavity fields take the form
\begin{align} \label{eq:steadyState}
    \ket{\alpha} & = (\mathcal{H} - (\Omega + \rmi \gamma))^{-1} \ket{f} \nonumber \\
    & = \sum_m \frac{\braket{\psi_m}{f}}{(E_m - \Omega) + \rmi \gamma} \ket{\psi_m},
\end{align}
where $\ket{\psi_m}$ are the eigenstates of $\mathcal{H}$ with corresponding eigenenergies $E_m$.

Eq.~\eqref{eq:steadyState} reflects the power and the limitations of the photonic quantum simulator introduced so far. It shows that the vector $\ket{\alpha}$ of coherent cavity fields corresponds, in steady state, to a superposition of eigenstates of the simulated Hamiltonian $\mathcal{H}$. Each eigenstate $\ket{\psi_m}$ contributes to the superposition according to (i) its spatial overlap $\braket{\psi_m}{f}$ with the vector $\ket{f}$ of driving-field amplitudes and (ii) its Lorentzian response to the driving frequency $\Omega$, with resonance frequency $E_m$ and linewidth $2\gamma$. Clearly, any eigenstate $\ket{\psi_m}$ can be addressed individually provided that the corresponding energy level $E_m$ is separated by more than half the linewidth from the rest of the spectrum. Eigenstates whose energy levels cannot be spectrally resolved (i.e., with an energy separation $\Delta \ll \gamma$), instead, contribute to Eq.~\eqref{eq:steadyState} with a relative weight solely determined by their spatial overlap with the driving field. This reflects one of the most important limitations of photonic quantum simulators and photonic systems in general; namely, the fact that energy bands cannot be filled uniformly as in fermionic systems where the Pauli exclusion principle prevails.

To circumvent the above limitations, we propose to simulate the bulk properties of an infinite system using a \emph{unit-cell} photonic quantum simulator. To this end, we assume that the non-interacting tight-binding model of interest (described by Eq.~\eqref{eq:genericHamiltonian}) is translationally invariant with respect to an enlarged unit cell of $N$ lattice sites consisting of $N_u$ sites in each spatial direction $u$~\cite{footnote:gaugeAndTranslationInvariance}. For simplicity, and without loss of generality, we focus on 2D models, in which case $u = 1, 2$ and $N = N_1 \times N_2$. Using Bloch's theorem, the infinite system that we wish to simulate can be described in momentum space by a $N \times N$ Hamiltonian $\mathcal{H}(\vect{k})$, and thus exhibits up to $N$ energy bands $E_n(\vect{k})$ with associated eigenstates $\ket{\psi_n(\vect{k})}$, where $\vect{k} = (k_1, k_2)$ with $k_u \in [-\pi / N_u, \pi / N_u]$ and $n$ is the index of the band. Crucially, one can equivalently simulate the Hamiltonian $\mathcal{H}(\vect{k})$ in a unit-cell array of $N$ coupled photonic cavities by encoding the momentum $\vect{k}$ in \emph{twisted} boundary conditions~\cite{Thouless82,Avron83}, i.e., periodic boundary conditions with an additional directional phase shift (or ``twist angle'') $\pm N_u \phi_u$ for tunneling across the boundary of the system in each direction $u$ [see Fig.~\ref{fig:unitCellSimulator}]. Physically, this stems from the fact that the phase $N_u \phi_u$ picked up by photons upon tunneling across the unit cell in a direction $u$ can be identified with the phase $N_u k_u$ accumulated, in an infinite system, by a plane wave $\rme^{\rmi \vect{k} \cdot \vect{r}}$ traveling across a unit cell in the same direction ($\vect{r}$ being the position vector). In the following, we shall therefore denote both the twist angles and the momentum by $\boldsymbol{\phi} = (\phi_1, \phi_2)$, i.e., $\boldsymbol{\phi} \equiv \vect{k}$.

In light of the above discussion, using a unit-cell photonic quantum simulator provides two major advantages: (i) Since each energy band generically reduces to a single eigenstate, a much higher spectral resolution can be achieved: Provided that $\mathcal{H}(\boldsymbol{\phi})$ is gapped and that the photon loss rate $\gamma$ is much smaller than the minimum gap, the frequency $\Omega$ of the driving field can be tuned into resonance with any targeted eigenstate $\ket{\psi_n(\boldsymbol{\phi})}$, giving rise to steady-state coherent cavity fields $\ket{\boldsymbol{\alpha}} \approx \ket{\psi_n(\boldsymbol{\phi})}$ [i.e., the sum in Eq.~\eqref{eq:steadyState} reduces to one term], up to an irrelevant complex factor~\cite{footnote:spatialOverlap}. In other words, any eigenstate $\ket{\psi_n(\boldsymbol{\phi})}$ of the simulated Hamiltonian $\mathcal{H}(\boldsymbol{\phi})$ can be directly \emph{observed} by measuring the amplitude and phase of the light emitted in steady state by each of the $N$ cavities of the unit-cell photonic lattice. We note that the spectrum $E_n(\boldsymbol{\phi})$ can also be measured using standard optical spectroscopy techniques---by monitoring, e.g., the total number of photons transmitted into the system as a function of the frequency $\Omega$ of the driving field. (ii) The ease with which the twist angles $\boldsymbol{\phi}$ can in principle be tuned in photonic systems---by modifying the optical path lengths coupling the cavities across the boundaries of the system---makes it possible to simulate the bulk properties of arbitrarily large systems~\cite{footnote:arbitrarilyLarge} using a much smaller (unit-cell) photonic lattice. Most importantly, it provides a way to simulate large non-interacting quantum systems---bosonic or fermionic~\cite{footnote:bosonicFermionic}---in an intrinsically \emph{momentum-resolved} manner.

We remark that the unit cell of interest must have at least $2$ lattice sites in each spatial direction (i.e., $N_u \geq 2$ for all $u$) in order for our scheme to work without modifications, since twist angles $\phi_u$ can only be introduced in the tunneling between \emph{distinct} sites. We emphasize, however, that this is not a fundamental constraint: If $N_u = 1$ in a particular direction $u$, one can always introduce the required twist angle $\phi_u$ by modulating, instead, the resonance frequencies $\omega_i$ of the cavities~\cite{supplementalMaterial,Kraus12_1,Kraus12_2,Verbin13,Kraus13}. In any case, tunable twisted boundary conditions and resonance frequencies can both be realized in photonic systems using standard techniques (see, e.g., Refs.~\cite{Kraus12_1,Hafezi13_2}).


\emph{Illustrative example: bulk topological properties of a quantum Hall-like system}---The unit-cell photonic quantum simulator introduced above provides a direct way to \emph{measure} both the spectrum and the wavefunctions of any targeted gapped non-interacting model with translation invariance, allowing to extract all of its bulk properties. Bulk topological properties, in particular, can be accessed through momentum-resolved measurements of the gauge-invariant spectral projection operator~\cite{Altland97,Schnyder08,Ryu10}
\begin{align} \label{eq:spectralProjector}
    P_n(\boldsymbol{\phi}) = \ket{\psi_n(\boldsymbol{\phi})} \bra{\psi_n(\boldsymbol{\phi})}.
\end{align}
This is to be contrasted to previous proposals allowing to probe topology in photonic lattices through the observation of edge states~\cite{Hafezi13_2}, approximate responses~\cite{Ozawa13}, or so-called ``Zak phases''~\cite{Longhi13}.

Below we demonstrate how to perform a unit-cell photonic simulation of the well-known Hofstadter model for 2D integer quantum Hall systems~\cite{Hofstadter76}. The Hofstadter Hamiltonian describes non-interacting particles hopping on a square lattice under a (real or synthetic) uniform perpendicular magnetic field and takes a similar form as the generic Hamiltonian of Eq.~\eqref{eq:genericHamiltonian}, namely,
\begin{align} \label{eq:quantumHallHamiltonian}
    H_{\rm Hofstadter} = t \sum_{\langle i,j \rangle} a^\dagger_i a_j \rme^{\rmi \phi_{ij}},
\end{align}
where the sum is restricted to nearest-neighboring sites, with a uniform hopping amplitude $t > 0$ and vanishing on-site potentials. Whereas the phases $\phi_{ij}$ are gauge-dependent quantities, the phase $\phi_p = \sum_p \phi_{ij}$ accumulated when hopping once around an elementary square (or ``plaquette'') of the lattice in the counter-clockwise direction is gauge-independent, and physically corresponds to an effective magnetic field. Here we assume that the number $n_p = (2\pi)^{-1} \phi_p$ of magnetic flux quanta per plaquette is constant and rational, i.e., $n_p = p/q$ with co-prime integers $p$ and $q$, so that the corresponding effective magnetic field is uniform and commensurate with the lattice. In that case the Hamiltonian is translationally invariant with respect to the so-called ``magnetic'' unit cell, which contains $q$ lattice sites~\cite{Thouless82,Avron83,Kohmoto85,Dana85}. More importantly, the infinite system to be simulated generically exhibits $q$ energy bands with non-trivial topological properties: Each band $n$ is characterized by an integer topological invariant known as the (first) Chern number~\cite{Thouless82,Avron83,Kohmoto85}, defined as
\begin{align} \label{eq:chernNumber}
    \nu_n = \frac{1}{2\pi \rmi} \int_{0}^{2\pi} \rmd \phi_1 \int_{0}^{2\pi} \rmd \phi_2 \, \Tr \left(P_n [\partial_{\phi_1} P_n, \partial_{\phi_2} P_n] \right),
\end{align}
where $P_n = P_n(\phi_1, \phi_2)$ is the spectral projection operator associated with the band of interest [cf. Eq.~\eqref{eq:spectralProjector}]. The Chern number is a topological property of an entire band which thus manifests itself most readily in fermionic systems where each filled band $n$ contributes to the overall Hall conductance by an integer value corresponding to its Chern number $\nu_n$ (in units of $e^{2}/h$)~\cite{Thouless82,Avron83,Kohmoto85}. In bosonic systems, however, determining the Chern number of specific bands via observable quantities presents a much greater challenge. Remarkably, this can be achieved rather straightforwardly using the unit-cell photonic quantum simulator introduced above, as we now proceed to demonstrate.

\begin{figure}[t]
    \begin{center}
        \includegraphics[width=\columnwidth]{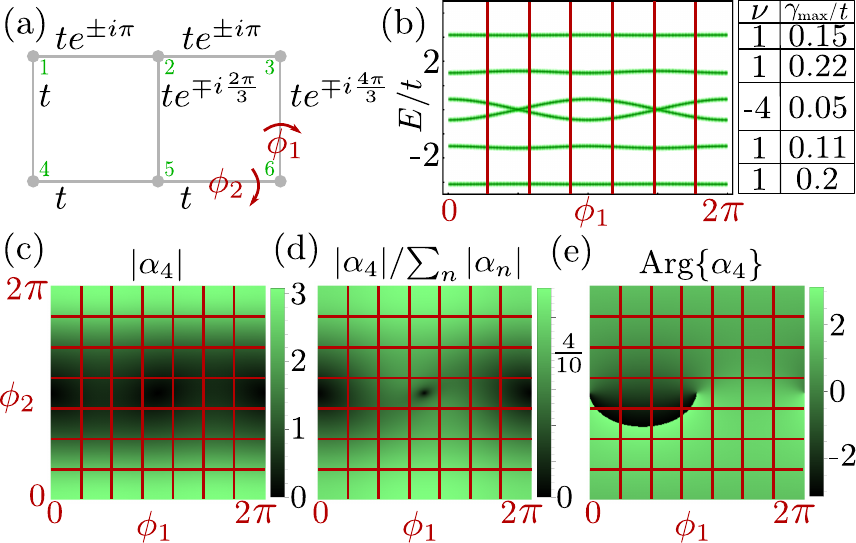}
        \caption{(Color online). Unit-cell photonic simulation of a 2D quantum Hall-like Hofstadter model with $n_p = p/q = 1/6$ effective magnetic flux quanta per (square-lattice) plaquette. (a) Relevant (magnetic) unit cell containing $q = 6$ lattice sites, with uniform hopping amplitudes $t$ and hopping phases $\phi_{ij}$ (or gauge; see Eq.~\eqref{eq:quantumHallHamiltonian}) chosen here so that $N_1 = 3$, $N_2 = 2$. Twisted boundary conditions are indicated by the corresponding twist angles $\phi_1$ and $\phi_2$. (b) Left: Spectrum $E(\phi_1, \phi_2)$ (shown here as a function of $\phi_1$ with $\phi_2 = 0$) exhibiting $q = 6$ energy levels which generate the bands of the simulated infinite system as the twist angles are continuously varied. Right: Table presenting the Chern number $\nu$ extracted for each band using a coarsely discretized ($7 \times 7$ grid) twist-angle space, along with the (estimated) maximum photon loss rate $\gamma_\mathrm{max}$ allowing to extract it. Note that $\gamma_\mathrm{max}$ is much smaller for the central band because the latter consists of $2$ energy levels which must be spectrally resolvable---for all chosen values of $\phi_1$ and $\phi_2$---for our scheme to work. (c) From left to right: Steady-state coherent-field amplitude, normalized amplitude, and phase found in cavity $4$ (see (a)), for example, as a function of the twist angles (also shown is the $7 \times 7$ discretization grid that is used). All three plots were obtained for $t = 1$ and $\gamma = 0.1$. The driving was applied to cavity $1$ with a coherent field amplitude $f = 1$ and frequency $\Omega$ tuned into resonance with the lowest energy level $E(\phi_1, \phi_2)$ (see (b)).}
        \label{fig:illustrativeExample}
    \end{center}
\end{figure}

To perform a unit-cell photonic simulation of the Hofstadter model, one must engineer a photonic lattice system described by the corresponding Hamiltonian [Eq.~\eqref{eq:quantumHallHamiltonian}]. More specifically, one must choose an appropriate gauge~\cite{footnote:gaugeInvariance} and restrict the system to a single magnetic unit cell with twisted boundary conditions and tunable twist angles $\phi_1$ and $\phi_2$~\cite{footnote:unitCellSizeRequirement}. Here we numerically investigate the case $n_p = 1/6$ and choose a gauge in which the (magnetic) unit cell has dimensions $3 \times 2$ [see Fig.~\ref{fig:illustrativeExample}(a)]. The unit-cell photonic system exhibits $6$ energy levels, as expected, and two of these levels form a single band as a function of the twist angles $(\phi_1, \phi_2)$ [see Fig.~\ref{fig:illustrativeExample}(b)]. Most importantly, the spectral projector $P_n(\phi_1, \phi_2)$ associated with a particular band $n$ can be \emph{measured} for all values of $\phi_1$ and $\phi_2$ at which the energy level(s) forming the band of interest can be spectrally resolved despite the broadening due to photon losses with rate $\gamma$. For bands formed by a single energy level and separated by a gap $\Delta$ from the rest of the spectrum, $\gamma \ll \Delta$ is the only requirement. Once measurements of $P_n(\phi_1, \phi_2)$ have been performed for a discrete set of values $\phi_1, \phi_2 \in [0, 2\pi)$, the Chern number $\nu_n$ of the corresponding band can be derived from its very definition, using a discretized version of Eq.~\eqref{eq:chernNumber}. We note that a coarse sampling of $P_n(\phi_1, \phi_2)$ over the ``Brillouin zone'' $(\phi_1, \phi_2)$ is often sufficient to determine the value of $\nu_n$ correctly~\cite{Fukui05}. We present in Fig.~\ref{fig:illustrativeExample}(b) a table showing the correct values of $\nu_n$ that would be measured for each of the bands using a $7 \times 7$ sampling grid, along with the maximum photon loss rate $\gamma_{\rm max}$ allowed for a correct measurement. Figs.~\ref{fig:illustrativeExample}(c)-(e) illustrate the typical data that must be obtained, for each cavity, to extract the Chern number of a specific band.

The fact that translation invariance is intrinsically preserved in a unit-cell simulation allows us to go beyond the above illustrative example. In particular, phases protected by point group or inversion symmetries are natural candidates to be studied with our scheme~\cite{Altland97,Schnyder08,Kitaev09,Ryu10,Fu11,Hsieh12,Tanaka12,Okada13}. In the absence of spatial symmetries, a total of ten symmetry classes can in principle be reached depending on the presence of time-reversal, particle-hole or chiral symmetry~\cite{Altland97}. Here, any spatial and non-spatial symmetry can be introduced by engineering the underlying photonic system. Photonic lattices with time-reversal symmetry, for example, have recently been realized~\cite{Jia13}.


\emph{Conclusion}---We have proposed a powerful practical scheme to simulate the bulk properties (topological, in particular) of arbitrary gapped non-interacting quantum systems of bosons or fermions in photonic lattices. We anticipate that our proposal will allow for the observation of bulk topological features that have so far only been theoretically predicted, such as topological phenomena protected by spatial symmetries~\cite{Fu11,Hsieh12,Tanaka12,Okada13} or entanglement spectra~\cite{Li08}. Beyond topology, we expect the concept of unit-cell photonic quantum simulator introduced in this work to prove useful in the investigation of a broad variety of bulk phenomena. Our proposal ultimately relies on two essential ingredients: (i) the restriction to small photonic lattice systems in order to achieve a high spectral resolution, and (ii) the use of tunable twisted boundary conditions to probe different momentum sectors. One important future direction will be to use similar ideas to study (small) photonic lattice systems with interactions.


The authors would like to thank M. Hafezi for motivating this endeavor, A. \.Imamo\u glu and G. Blatter for their support,  and G.-M. Graf, A. Srivastava, and E. van Nieuwenburg for fruitful discussions. Financial support from the Swiss National Science Foundation (SNSF) is gratefully acknowledged.


\bibliographystyle{apsrev}
\bibliography{manuscript_bibliography}

\begin{thebibliography}{40}
\bibitem[{foo({\natexlab{a}})}]{footnote:gaugeAndTranslationInvariance}
\bibinfo{note}{The invariance of the Hamiltonian under translations is
  gauge-dependent since the phases $\phi_{ij}$ are gauge-dependent quantities.
  Here we always assume that a proper gauge has been chosen.}
\bibitem[{foo({\natexlab{b}})}]{footnote:spatialOverlap}
\bibinfo{note}{We assume, without loss of generality, that the spatial overlap
  $\braket{\psi_n(\boldsymbol{\phi})}{f}$ between the targeted eigenstate and
  the vector of driving-field amplitudes is non-negligible.}
\bibitem[{foo({\natexlab{c}})}]{footnote:arbitrarilyLarge}
\bibinfo{note}{The maximum system size that can be simulated in the unit cell
  is limited by (i) the resolution that can be achieved for varying the twist
  angles, and (ii) the broadening of the energy levels due to photon losses.}
\bibitem[{foo({\natexlab{d}})}]{footnote:bosonicFermionic}
\bibinfo{note}{Since the unit-cell photonic system simulates first-quantized
  single-particle Hamiltonians, the bosonic or fermionic nature of the
  underlying non-interacting quantum system is irrelevant.}
\bibitem[{foo({\natexlab{e}})}]{footnote:gaugeInvariance}
\bibinfo{note}{The gauge is determined by the physical implementation of the
  unit-cell photonic simulator, since gauge invariance is broken by dissipation
  (photon losses).}
\bibitem[{foo({\natexlab{f}})}]{footnote:unitCellSizeRequirement}
\bibinfo{note}{We recall that $N_1, N_2 \geq 2$ is required in order to
  introduce the required twist angles $\phi_1$ and $\phi_2$ associated with
  each spatial dimension without having to modulate on-site potentials. Here
  the numbers $N_1$ and $N_2$ defining the size of the unit cell can be
  modified at will by a suitable choice of gauge (i.e., of phases $\phi_{ij}$
  in Eq.~\eqref{eq:quantumHallHamiltonian}) as long as $N_1 \times N_2 = q$.}
\end{thebibliography}

\end{document}